\documentclass[%
  pop,superscriptaddress,
 amsmath,amssymb,
 preprint,%
]{revtex4-1}

\usepackage{CJK}
\usepackage{graphicx}
\usepackage{dcolumn}
\usepackage{bm}

\usepackage[utf8]{inputenc}
\usepackage[T1]{fontenc}
\usepackage{mathptmx}
\usepackage{etoolbox}
\usepackage{booktabs}
\makeatletter
\def\@email#1#2{%
 \endgroup
 \patchcmd{\titleblock@produce}
  {\frontmatter@RRAPformat}
  {\frontmatter@RRAPformat{\produce@RRAP{*#1\href{mailto:#2}{#2}}}\frontmatter@RRAPformat}
  {}{}
}%
\makeatother
\begin{document}
\begin{CJK*}{UTF8}{}
\title[Prediction of Mode Structure Using A Novel Physics-Embedded Neural ODE Method]{Prediction of Mode Structure \\ Using A Novel Physics-Embedded Neural ODE Method}
\author{Bowen ZHU \CJKfamily{gbsn}(朱~博文)}
\affiliation{ 
School of Electrical Engineering, Xi'an Jiaotong University,
Xi'an, Shaanxi, China
}
\author{Hao WANG \CJKfamily{gbsn}(王~灏)}%
\affiliation{ 
National Institute for Fusion Science, National Institutes of Natural Sciences, Toki, Japan
}
 \email{wanghao@nifs.ac.jp}

\author{Jian WU \CJKfamily{gbsn}(吴~坚)}
\affiliation{ 
School of Electrical Engineering, Xi'an Jiaotong University,
Xi'an, Shaanxi, China
}

\author{Haijun REN \CJKfamily{gbsn}(任~海骏)}
\affiliation{%
CAS Key Laboratory of Geospace Environment and Department of Engineering and Applied Physics, School of Physical Sciences, University of Science and Technology of China, Hefei, China
}%

\date{\today}

\begin{abstract}
We designed a new artificial neural network by modifying the neural ordinary differential equation (NODE) framework to successfully predict the time evolution of the 2-dimensional mode profile in both the linear growth stage and nonlinear saturated stage.
Starting from the magnetohydrodynamic (MHD) equations, simplifying assumptions were applied based on physical properties and symmetry considerations of the energetic-particle-driven geodesic acoustic mode (EGAM) to reduce complexity.
Our approach embeds known physical laws directly into the neural network architecture by exposing latent differential states, enabling the model to capture complex features in the nonlinear saturated stage that are difficult to describe analytically, and thus, the new artificial neural network is named as ExpNODE (Exposed latent state Neural ODE).
ExpNODE was evaluated using a data set generated from first-principles simulations of the EGAM instability, focusing on the pre-saturated stage and the nonlinear saturated stage where the mode properties are most complex.
Compared to state-of-the-art models such as ConvLSTM, ExpNODE with physical information not only achieved lower test loss but also converged faster during training.
Specifically, it outperformed ConvLSTM method in both the 20-step and 40-step prediction horizons, demonstrating superior accuracy and efficiency.
Additionally, the model exhibited strong generalization capabilities, accurately predicting mode profiles outside the training data set.
Visual comparisons between model predictions and ground truth data showed that ExpNODE with physical information closely captured detailed features and asymmetries inherent in the EGAM dynamics that were not adequately captured by other models.
These results suggest that integrating physical knowledge into neural ODE frameworks enhances their performance, and provides a powerful tool for modeling complex plasma phenomena.
\end{abstract}

\maketitle
\end{CJK*}

\section{Introduction}\label{sec1}

In tokamak and stellarator devices, the behavior of magnetohydrodynamic (MHD) modes significantly affects plasma confinement and stability. Among these modes, geodesic acoustic modes (GAMs)\citep{Winsor.1968,Diamond} and their energetic-particle-driven counterparts, energetic-particle-driven geodesic acoustic modes (EGAMs)\citep{fu2008energetic,Nazikian.2008,Wang.2013.PRL,Zarzoso.2013,Sasaki.2017,Wang.PRL2,Wang.2019,conway2021geodesic}, are of particular interest due to their impact on energetic particle transport, on anomalous bulk ion heating, and on plasma turbulence.

EGAMs are oscillations in the plasma caused by the interaction between energetic particles and the bulk plasma, often resulting from neutral beam injection \citep{fu2008energetic,conway2021geodesic}. Understanding and predicting EGAM behavior is essential for optimizing plasma confinement. However, the simulation of EGAMs poses significant challenges of computing resources due to the complex interactions between energetic particles and the bulk plasma.

First-principles simulations of EGAMs rely on solving the full set of governing equations, such as the MHD equations or the kinetic equations \citep{Wang.2013.PRL,Zarzoso.2013,Biancalani.2017,Wang.2019,Rettino.2022}. While these simulations can capture detailed physics, they are computationally intensive.

In recent years, machine learning (ML) techniques have emerged as powerful tools for modeling complex physical systems, offering the potential to learn patterns and dynamics directly from data \citep{ling2016reynolds}. Deep learning models, such as convolutional neural networks (CNNs) and recurrent neural networks (RNNs), have been applied to various problems in plasma physics, including disruption prediction \citep{ML.disruption,ML.disruption24}, plasma control \citep{ML.TCV,ML.control}, turbulence understanding \citep{ML.blob}, and identification of experimental conditions \citep{ML.laser}. However, purely data-driven models may lack generalizability and fail to capture essential physical constraints, leading to limited applicability outside the training data domain.

To address these challenges, integrating physical knowledge into ML models has gained attention as a means to improve their accuracy and robustness \citep{karniadakis2021physics}. Physics-informed neural networks (PINNs) enforce physical laws by embedding differential equations into the learning process, guiding the model toward physically consistent solutions \citep{raissi2019physics}. This approach combines the strengths of data-driven models and traditional physics-based methods, enabling the modeling of complex systems where full analytical solutions are intractable.

In this work, we propose a modified neural ordinary differential equation (Neural ODE \citep{chen2018neural}) framework, referred to as ExpNODE (Exposed latent state Neural ODE), which is designed to model the dynamics of EGAMs. Also, by embedding simplified physical laws and symmetries of EGAM into ExpNODE, we obtained an enhanced model named ExpNODE-p (in other words, ExpNODE with physics). By exposing the latent states, our framework departs from the traditional Neural ODE approach, where the latent states are governed by a differential flow. Our method begins by simplifying the governing MHD equations using physical insights and symmetry considerations specific to toroidally confined plasmas. By making reasonable assumptions about the dominance of certain terms and exploiting the symmetry properties of the system, the complexity of the equations is reduced while retaining the critical dynamics of EGAMs.

We evaluate the proposed method using data generated from first-principles simulations of EGAM instabilities, comparing its performance against state-of-the-art models such as ConvLSTM and ExpNODE. Our results demonstrate that the ExpNODE-p model not only achieves higher accuracy in predicting the complex dynamics of EGAMs but also converges faster during training. Additionally, the model exhibits strong generalization capabilities, accurately predicting system behavior outside the training data range.

The rest of the paper is organized as follows. Section~\ref{sec2} presents the method, including the simplification of the governing MHD equations and the development of the ExpNODE-p framework. Section~\ref{sec3} details the results of our experiments, comparing the performance of our model with other approaches. Section ~\ref{sec4} discusses the implications of our findings and the applicability of the simplification methods and modeling approach to other plasma physics problems. Finally, section~\ref{sec5} concludes the paper and outlines directions for future work.

\section{Method}\label{sec2}

The EGAM data set in the present work was generated in Wang's simulation\citep{Wang.2013.PRL,Wang.PoP}, where a first-principles code, MEGA\citep{Todo.1998,Todo.2006}, is applied and EGAM dynamics are governed by the MHD equations and the drift-kinetic equations. These equations are used in cylindrical coordinate ($R$, $\phi$, $z$) to describe the bulk plasma and the energetic particles. Our goal is not to accurately reproduce the simulation results of the first-principles code by machine learning method, but to quickly predict the physical results by providing the dominant physical features to the machine learning model. Thus, we focus on the most important physical quantities for mode profile evolution, but ignore all the other physics.
The dominant poloidal mode numbers of the EGAM are \(m = 0\) and \(m = 1\), where \(m\) represents the poloidal mode number\citep{Winsor.1968,Diamond,Wang.PoP,conway2021geodesic}. Therefore, we will focus on predicting only the physical quantities that strongly affect the \(m = 0\) and \(m = 1\) components, such as the electric field and density.
Although the magnetic perturbation that strongly affects the \(m = 2\) component was also extensively investigated due to its physical significance\citep{Ren.2014,Huang.2019,conway2021geodesic,YHWang.2024}, its strength is much weaker than that of the \(m = 0\) and \(m = 1\) components.
If both the very weak magnetic perturbation and the very strong density perturbation are used simultaneously in machine learning, the weight of the weaker perturbation in the neural network could be disproportionately low, potentially degrading network performance. For this reason, the \(m = 2\) component is excluded from consideration in the present paper.
Also, we focus on the 2-dimensional mode profiles and thus \(\partial/\partial \phi\) is not considered.

\subsection{The dominant physical quantity for \(m = 0\): Electric Field}
Both the electric field $\mathbf{E}$ and velocity $\mathbf{v}$ are important for \(m = 0\) perturbations. However, velocity is very different between high field side and low field side, in other words, the symmetry of velocity is not as good as electric field. The present research utilized symmetry to enhance the ability of the neural network, and thus, electric field is a better candidate for mode profile prediction and we do not predict the profile of velocity.

The electric field is decided by Ohm's law:
\begin{align}
    E_R &= -v_{\phi} B_z + v_z B_{\phi}, \label{eq:ER_relation} \\
    E_z &= -v_R B_{\phi} + v_{\phi} B_R, \label{eq:EZ_relation} \\
    E_{\phi} &= -v_z B_R + v_R B_z \approx 0. \label{eq:Ephi_relation}
\end{align}
Based on the assumption that electrons have no mass, the parallel electric field is zero, and $E_\phi$ is also negligible.
Assuming \( B_R, B_z \ll B_{\phi} \) in the EGAM region:
\begin{equation}
    v_R = -\frac{E_z}{B_{\phi}}, \quad v_z = \frac{E_R}{B_{\phi}}. \label{eq:velocity_components}
\end{equation}
Defining the electric field in the poloidal cross section \( E_{\text{pol}} = \sqrt{E_R^2 + E_z^2} \), and using (\ref{eq:Ephi_relation}) and (\ref{eq:velocity_components}), we express it as:

\begin{equation}
    E_{\text{pol}} = K_{E}(R, z) |E_z|, \quad \text{with} \quad K_{E}(R, z) = \sqrt{1 + \left( \frac{B_z}{B_R} \right)^2}. \label{eq:Epol_K}
\end{equation}
In terms of the properties of the EGAM, the magnetic perturbation is an order of magnitude or more smaller than the dominant perturbation, and thus ${\partial \mathbf{B}}/{\partial t} \approx 0$. 
Then, here \( K_{E}(R, z) \) is time-invariant and symmetric with respect to \( z \).

Centering at the magnetic axis where \( R = R_{\text{0}} \), the system exhibits:
\begin{enumerate}
    \item \textbf{Antisymmetry of \( E_z \)}:
    \( E_z(R, z) = -E_z(R, -z)\).
    \item \textbf{Poloidal symmetry of \( E_{\text{pol}} \)}:
      \( E_{\text{pol}}(R_1, z_1) = E_{\text{pol}}(R_2, z_2)\) if \(\sqrt{(R_1-R_0)^2 + z_1^2} = \sqrt{(R_2-R_0)^2 + z_2^2}\).
      It can be re-written as \( E_{\text{pol}}(R, z) = E_{\text{pol}}(\rho)\), where \(\rho = \sqrt{(R-R_0)^2 + z^2}\).
    \item \textbf{Up-Down Symmetry of \( K_E \)}:
    \( K_E(R, z) = K_E(R, -z). \)
\end{enumerate}

To incorporate these symmetries, we define:
\begin{equation}
    K_{\text{mod}}(R, z) = \begin{cases}
    K_E(R, z), & z \geq 0, \\
    -K_E(R, z), & z < 0.
    \end{cases} \label{eq:K_mod}
\end{equation}
Thus, the modified poloidal electric field is defined as:
\begin{equation}
    E_{\text{pol, mod}}(R, z) = K_{\text{mod}}(R, z) \cdot E_z(R, z) = E_{\text{pol, mod}}^{1D}(\rho), \label{eq:Epol_mod}
\end{equation}
which depends solely on \( \rho \), resulting in a one-dimensional profile. Consequently, \( E_z \) can be reconstructed from \( E_{\text{pol, mod}}^{1D}(\rho) \) as follows:
\begin{equation}
    E_z(R, z, t) = \frac{E_{\text{pol, mod}}^{1D}(\rho,t)}{K_{\text{mod}}(R, z)}. \label{eq:Ez_reconstructed}
\end{equation}
Similarly, the time derivative of \( E_z \) can be obtained in the same manner:
\begin{equation}
    \frac{dE_z}{dt} = \frac{E_{\text{pol, mod}}^{1D}(\rho,t)'}{K_{\text{mod}}(R, z)} = \frac{L(\rho,t)}{K_{\text{mod}}(R, z)}, \label{eq:derivative_Ez_reconstructed}
\end{equation}
where
\[
L(\rho,t) = E_{\text{pol, mod}}^{1D}(\rho)'.
\]
This shows that \( E_z \) and its time derivative can be derived from a latent one-dimensional function, which we will explore in the following section.

\subsection{The dominant physical quantity for \(m = 1\): density}
Both the density $n$ and pressure $p$ are important for \(m = 1\) perturbations. Since density and pressure perturbations are in phase, here, we only focus on the density but do not predict the pressure perturbation.

Assuming small perturbations around equilibrium (\(n = n_0 + n_1,\, \mathbf{v} = \mathbf{v}_0 + \mathbf{v}_1 = \mathbf{v}_1 \)) and neglecting second-order and advection terms, we linearize the continuity equations:
\begin{equation}
    \frac{\partial n}{\partial t} = \frac{\partial n_1}{\partial t} = -n_0 \left( \frac{v_R}{R} + \frac{\partial v_R}{\partial R} + \frac{\partial v_z}{\partial z} \right), \label{eq:rho_lin}
\end{equation}
According to Faraday's law and ${\partial \mathbf{B}}/{\partial t} \approx 0$ assumption, we have
\begin{equation}
    \frac{\partial E_R}{\partial z} = \frac{\partial E_z}{\partial r}. \label{eq:ER_EZ_relation}
\end{equation}
Substituting \( v_R \) and \( v_z \) from (\ref{eq:velocity_components}) and using (\ref{eq:ER_EZ_relation}), the linearized continuity equation simplifies to:
\begin{equation}
    \frac{\partial n}{\partial t} = \frac{\partial n_1}{\partial t} = n_0 \left( \frac{1}{R B_{\phi}} - \frac{1}{B_{\phi}^2} \frac{\partial B_{\phi}}{\partial R}  \right) E_z, \label{eq:rho_simplified}
\end{equation}
where $B_\phi = B_{\phi 0}+B_{\phi 1} = B_{\phi 0}$, $E_z = E_{z0} + E_{z1} = E_{z1}$.
We can precompute the factor multiplying \( E_z \):
\begin{equation}
    \text{K}_\rho(R,z) = \frac{1}{R B_{\phi}} - \frac{1}{B_{\phi}^2} \frac{\partial B_{\phi}}{\partial R}, \label{eq:rho_factor}
\end{equation}
and this factor will be embedded in the neural network.

\subsection{Modified Neural ODE Framework for Physical System Modeling}
Appropriate physical information can improve the convergence speed and prediction accuracy of neural network, but too much or too complicated physical information will increase the workload of neural network and degrade the prediction ability. We finally selected only a few simple physical information to predict $E_z$ and $n$.
Then, building upon Ref.~\citep{INODE} on neural ODE, our proposed ExpNODE-p model is specifically designed for physical systems, incorporating time-series image data and directly exposing latent differential states.
The streamlined design and enhanced interpretability of ExpNODE-p make it a powerful tool for solving complex physical problems. By integrating known physical laws with data-driven learning, our approach captures complex dynamics that are challenging to model analytically. This methodology diverges from conventional neural ODE paradigms by embedding partial physical knowledge and enforcing symmetry constraints.

Our method combines CNN and multi-layer perceptrons (MLPs) within a neural ODE framework to model the system's dynamics. Our approach differs from the vanilla Neural ODE \citep{chen2018neural} in several ways:
The model handles  two-dimensional time-series data instead of one-dimensional trajectory data.
Rather than using the regular encoder-decoder setup and allowing the differential flow in the latent space, we expose the differential flow as the output of the model, enabling easy manipulation of derivatives.

The approach proceeds as follows (see Table~\ref{tab:model_architecture_compact} for detailed architecture)::
\begin{enumerate}
    \item \textbf{Input Representation}: Represent \( E_z \) and \( n \) as separate channels in a multi-channel input tensor (image).

    \item \textbf{Initial Linear Computation for \( n \)}: Compute an initial estimate of \( {\partial n}/{\partial t} \) using the known linear relationship in Equation~\eqref{eq:rho_simplified}.

    \item \textbf{Shared Feature Extraction}: Process the multi-channel input through shared convolutional layers to extract common spatial features and capture interactions between \( E_z \) and \( n \).

    \item \textbf{Branching into Specialized Paths}:
    \begin{itemize}
        \item \textbf{\( E_z \) Path}: Further process shared features through convolutional and fully connected layers specific to \( E_z \), encoding it into a latent representation L($\rho$,t). In practice, the latent state is represented by a set of Fourier coefficients. To meet the specific requirements of the radial profile where both the normalized value and its derivative are zero at the last closed flux surface, we define:
          $$L(\rho,t)=(1+\rho/a)^2 L_{Fourier}(\rho,t)$$
          where $a$ is the minor radius. Decode this latent state, as per Equation~\eqref{eq:derivative_Ez_reconstructed}, to compute \( {\partial E_z}/{\partial t} \), incorporating nonlinear dynamics not fully captured by the linear model and enforcing symmetry.
          
        \item \textbf{\( n \) Path}: Process shared features through layers specific to \( n \). Combine the network output with the initial linear computation to update \( \frac{\partial n}{\partial t} \) with learned nonlinear contributions.
    \end{itemize}
\end{enumerate}



\begin{table}[h]
    \centering
    \caption{ExpNODE-p  architecture. The "hidden size" is a tunable parameter, $N_{terms}$ represents the number of terms in the Fourier approximation, "Conv" is the abbreviation of convolution, and "FC" is the abbreviation of fully connected.}
    \begin{tabular}{lll}
        \toprule
        \textbf{Component \quad \quad \quad} & \textbf{Layer Type \quad \quad \quad} & \textbf{Size \quad \quad \quad} \\

        \midrule
        \textbf{Shared Feature } & Conv & hidden size \\
          \textbf{Extraction}     & Conv& $2 \times$ hidden size \\

        \midrule
        \textbf{$E_z$ Path} & Conv & hidden size \\
                           & Conv & 8 \\
                           & FC & 64 \\
                          & FC & 2$\times N_{terms}$-1 \\
        \midrule
        \textbf{$n$ Path} & Conv& $2 \times$ hidden size \\
                           & Conv & hidden size \\
                            & Conv& 1 \\
        \bottomrule
    \end{tabular}
    \label{tab:model_architecture_compact}
\end{table}
This architecture allows the integration of known linear dynamics with learned nonlinear effects, effectively capturing complex interactions and symmetries between \( E_z \) and \( n \).

The system's dynamics are expressed as:
\begin{equation}
    \frac{d}{dt}
    \begin{bmatrix}
    E_z \\
    n 
    \end{bmatrix}
    =
    \underbrace{
    \begin{bmatrix}
    0 & 0\\
    n_0 \cdot K_\rho(r) & 0 
    \end{bmatrix}
    \begin{bmatrix}
    E_z \\
    n
    \end{bmatrix}
    }_{\text{Known Linear Dynamics}}
    +
    \underbrace{
    \begin{bmatrix}
    \text{Decoder} & 0\\
    0 & I 
    \end{bmatrix}
    f_{\text{NN}}
    \left(
    \begin{bmatrix}
    E_z \\
    n
    \end{bmatrix}
    \right)}_{\text{Learned Nonlinear Dynamics}},
    \label{eq:combined_dynamics}
\end{equation}
where \( f_{\text{NN}} \) represents the nonlinear functions learned by the neural network, \( I \) is the identity matrix, and the \(\text{Decoder}\) transforms the latent representation back to the physical space, ensuring symmetry constraints are satisfied.

\begin{figure}
  \centering
  \includegraphics[width=1\columnwidth]{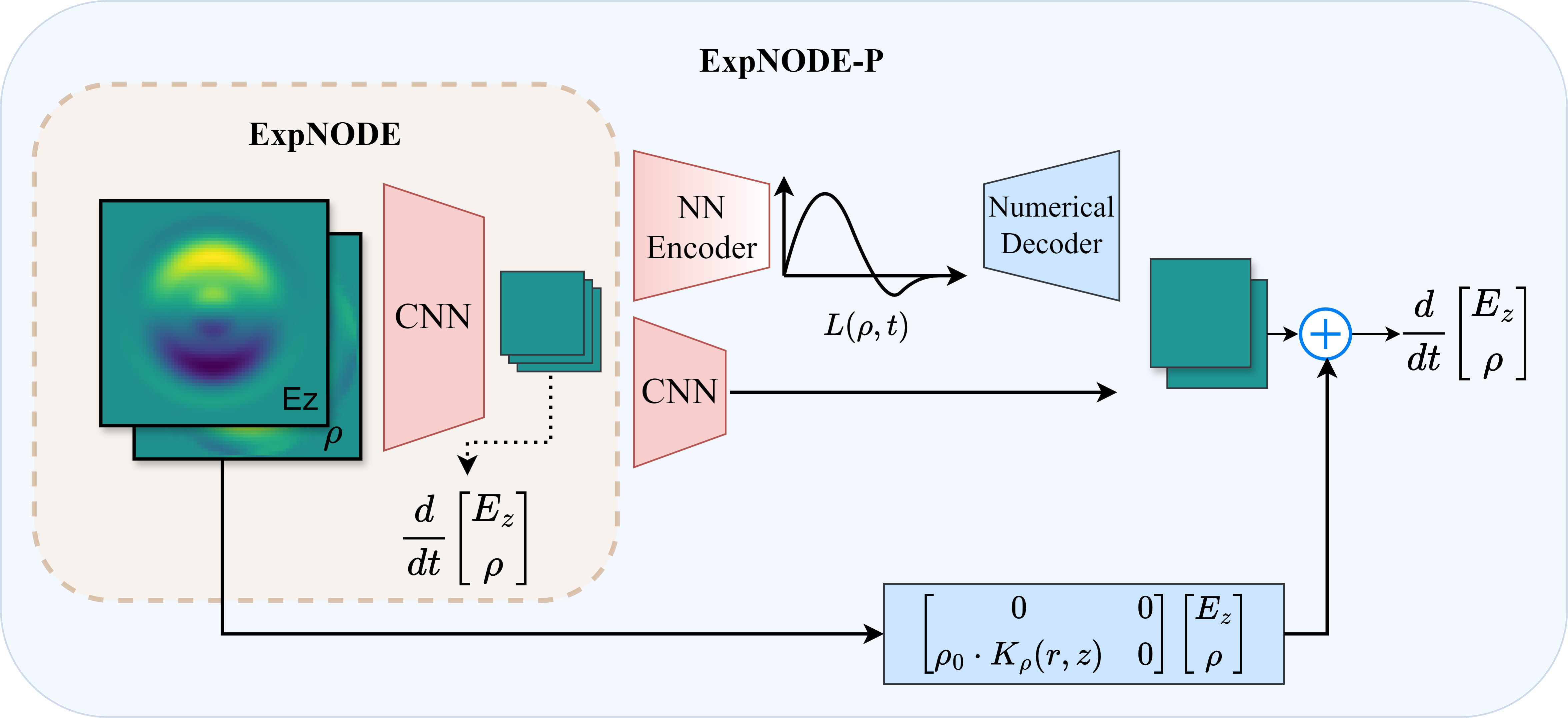}
\caption{Architecture of the modified neural ODE network. The orange dashed box highlights the ExpNODE architecture, where the differential state is directly output after the CNN layers. In the ExpNODE-p, the input consists of \( E_z \) and \( n \),  which are processed through shared convolutional layers before branching into specialized paths for each variable. These paths integrate known linear dynamics while learning the nonlinear contributions.}
  \label{fig:NN}
\end{figure}
The aforementioned neural network is illustrated in Fig.~\ref{fig:NN}.
It is important to note that due to the structure of our modified neural ODE, the physical information controlling the system can be entirely learned from data and applied to new scenarios. However, this approach performs worse and converges more slowly compared to when physical information is explicitly provided.

\section{Results}\label{sec3}

The ExpNODE-p framework was evaluated on a data set generated from MEGA simulations of the EGAM instability. The data set comprises 1,125 time steps of two-dimensional data, each represented as a multi-channel matrix. Each recorded time step corresponds to every 2,000 numerical simulation steps in the MEGA simulation.

In the EGAM data set of the present work, the first 850 time steps correspond to the linear stage, which is characterized by an exponential growth in the magnitude of perturbations while the perturbation patterns and frequencies remain unchanged. Predicting the mode profile in the linear stage is relatively straightforward, and conventional neural network methods have already achieved satisfactory performance in this stage. Therefore, our tests and comparisons focus on the pre-saturated and nonlinear stages (time steps 850 onwards), where the system exhibits more complex dynamics.

For our experiments, time steps 850 to 1,080 (230 steps) were used as the training data set, while the remaining time steps served as the test data set. To demonstrate the effectiveness of the proposed model, we compared it with state-of-the-art methods. Specifically, the following models were evaluated:

\textbf{ConvLSTM}: A Convolutional Long Short-Term Memory network, a leading model for spatiotemporal time-series prediction(without embedded physical information).

\textbf{ExpNODE}: The proposed framework without embedded physical information.

\textbf{ExpNODE-p}: The proposed framework with embedded physical information, as described in Section~\ref{sec2}.

All data were normalized prior to training. Although different models may be trained with different loss functions(See Table \ref{tab:hyperparameters}), the Mean Squared Error (MSE) was consistently used as the testing metric across all models to ensure a fair comparison.
\subsection{Model Training and Evaluation}
\begin{table}
  \centering
  \caption{Hyperparameters used for the models. Here, Adam (adaptive moment estimation), RMSprop (root mean square propagation), and LAE (least absolute errors) are optimization algorithms commonly used in training neural networks.}
  \label{tab:hyperparameters}
  \begin{tabular}{lccc}
    \toprule
    \textbf{Model} & \textbf{Params} & \textbf{Optimizer} & \textbf{Loss Function} \\
    \midrule
    \multicolumn{4}{c}{\textbf{20-step Prediction}} \\
    \midrule
    ExpNODE & 380k & Adam (lr=0.002) & MSE Loss \\
    ExpNODE-p & 380k & Adam (lr=0.002) & MSE Loss \\
    ConvLSTM & 450k & Adam (lr=0.001) & MSE Loss \\
    \midrule
    \multicolumn{4}{c}{\textbf{40-step Prediction}} \\
    \midrule
    ExpNODE & 100k & RMSprop (lr=0.0012) & LAE Loss \\
    ExpNODE-p & 100k & RMSprop (lr=0.0012) & LAE Loss \\
    ConvLSTM & 450k & Adam (lr=0.001) & MSE Loss \\
    \bottomrule
  \end{tabular}
\end{table}

We evaluated the models on two prediction horizons:

\textbf{20-step Prediction}: Models trained to predict the mode profiles in 20 future time steps, with testing from time steps 1,080 to 1,100.

\textbf{40-step Prediction}: Models trained to predict the mode profiles in 40 future time steps, with testing from time steps 1,080 to 1,120.

Although both models share the same training data set, they are trained differently due to their sequence lengths corresponding to the prediction horizon (20 or 40 steps). The sequence length used during training significantly influences the model's ability to capture dependencies over different temporal scales. Consequently, a model optimized for long-term prediction may not necessarily excel in short-term prediction, and vice versa.

Training time is defined as the time taken for the model's test loss to reach a plateau or the wall time for 2,000 epochs, whichever occurs first. The number of epochs at which training time is recorded is indicated. The final test error for ExpNODE-p includes post-processing where a Gaussian filter is applied to smooth the rough edges resulting from numerical decoding.
All training was performed using an NVIDIA RTX 2060 Max-Q GPU. The hyperparameters used for the models are summarized in Table~\ref{tab:hyperparameters}. The ExpNODE and ExpNODE-p were trained with the same hyperparameters within each prediction horizon to ensure a fair comparison.

\subsection{Model Performance}

\begin{figure} 
\centering 
\includegraphics[width=1\columnwidth]{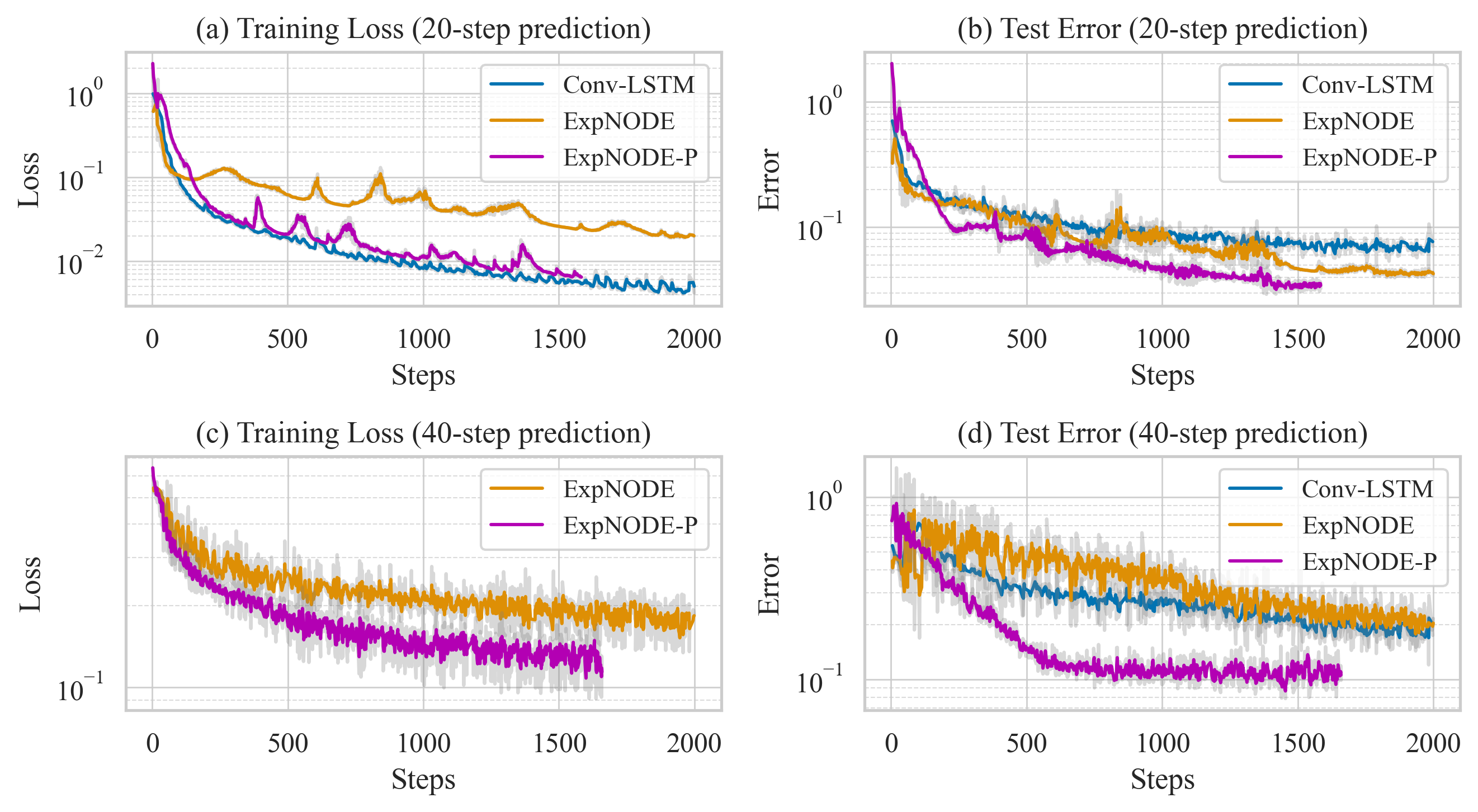} 
\caption{Training and test loss curves for the 20-step and 40-step prediction cases, showing the performance of ConvLSTM, ExpNODE, and ExpNODE-p for both prediction horizons. Note that in Figure (c), the ConvLSTM model is not included because it was trained using MSE Loss, whereas the ExpNODE and ExpNODE-p models were trained using LAE Loss; thus, their loss values are not directly comparable. Additionally, the same ConvLSTM model was used for both the 20-step and 40-step predictions.} \label{fig:train_val_curves} 
\end{figure}

Figure~\ref{fig:train_val_curves} shows the training and test loss curves for the models on both prediction horizons. Table~\ref{tab:model_comparison} presents the test loss and training time for each model and prediction horizon.

\begin{table}[h]
    \centering
    \caption{Comparison of model performance and training time for 20-step and 40-step predictions.}
    \label{tab:model_comparison}
    \begin{tabular}{lcc}
        \toprule
        \textbf{Model} & \textbf{Test Loss} & \textbf{Training Time} \\
        \midrule
        \multicolumn{3}{c}{\textbf{20-step Prediction}} \\
        \midrule
        ConvLSTM & 0.0688 & 55.2 min/2,000 epochs \\
        ExpNODE & 0.0411 & 45.6 min/2,000 epochs \\
        ExpNODE-p  & 0.0219 & 20.2 min/800 epochs \\
        \midrule
        \multicolumn{3}{c}{\textbf{40-step Prediction}} \\
        \midrule
        ConvLSTM & 0.2053 & 55.2 min/2,000 epochs \\
        ExpNODE & 0.2208 & 66.6 min/2,000 epochs \\
        ExpNODE-p & 0.08772 & 67.2 min/1,500 epochs \\
        \bottomrule
    \end{tabular}
\end{table}
From the results, we observe that the ExpNODE-p not only achieves the lowest test loss but also converges faster than the other methods. In the 20-step prediction case, the ExpNODE-p achieves a test loss of 0.02187, outperforming both the ConvLSTM and the ExpNODE. Similarly, in the 40-step prediction case, the ExpNODE-p demonstrates superior performance with a test loss of 0.08772, significantly lower than that of ConvLSTM and ExpNODE.

Moreover, the ExpNODE-p converges faster, requiring fewer epochs to reach a plateau in test loss. For the 20-step prediction, it converged within 800 epochs, compared to 2,000 epochs for the other models, resulting in shorter training times.

These results indicate that embedding physical information into the neural ODE framework enhances both the accuracy and efficiency of the model, enabling it to better capture the complex dynamics of the EGAM instability, particularly in the nonlinear regime.
\begin{figure}
\centering
\begin{tabular}{cc}
\includegraphics[width=0.45\columnwidth]{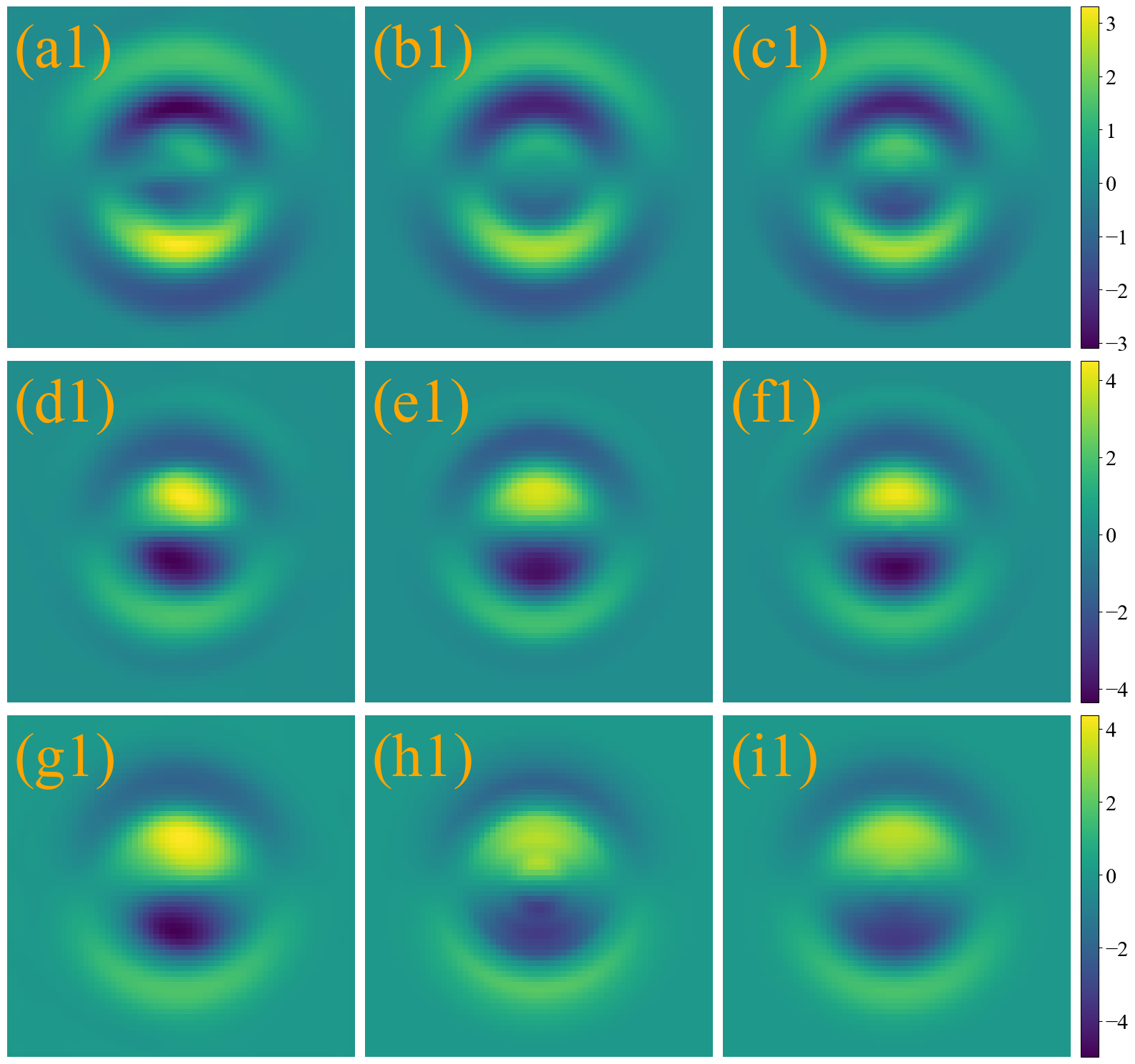} &
\includegraphics[width=0.45\columnwidth]{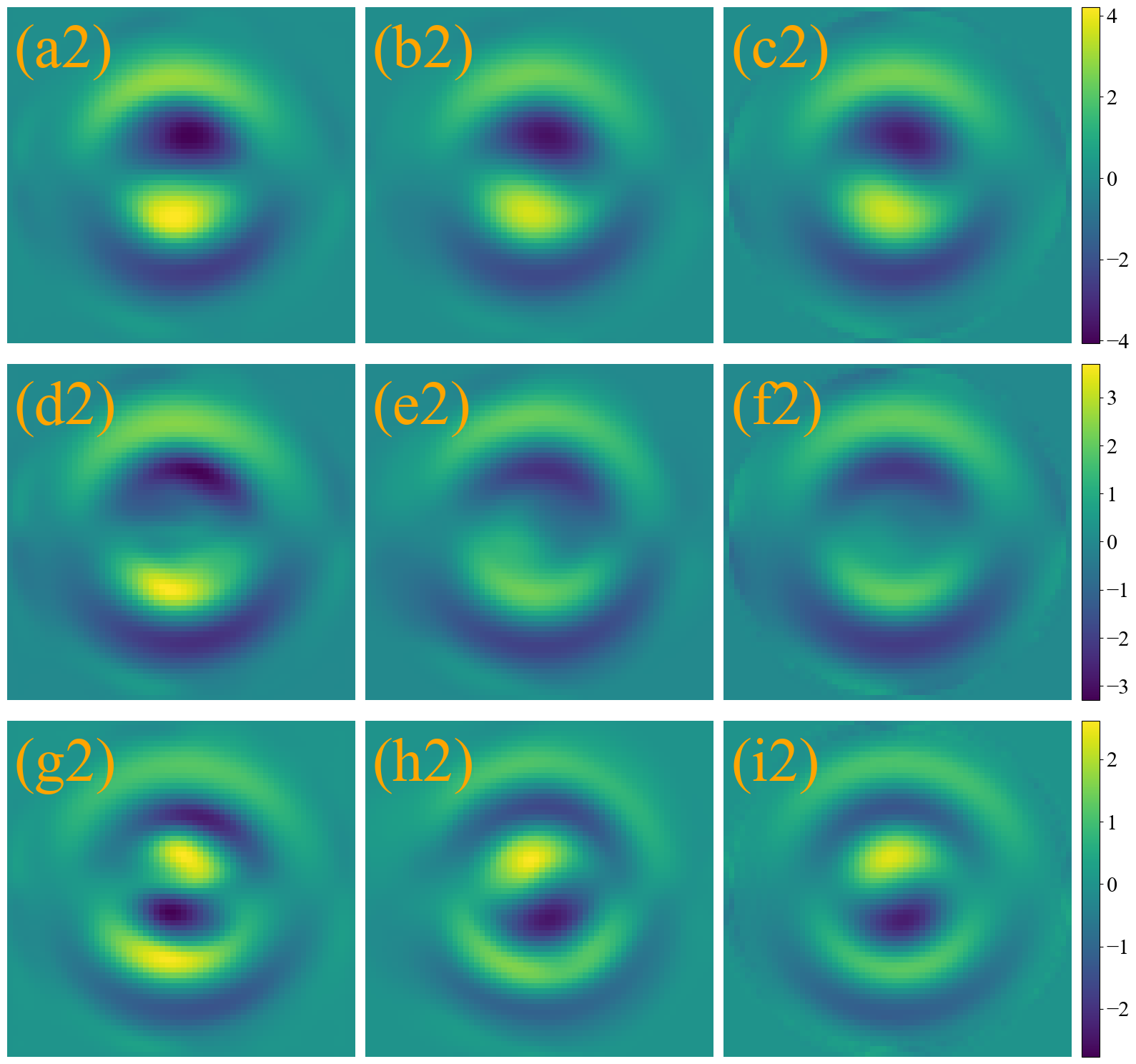} \\
(a) $E_z$ at steps 5, 10, and 15&
(b) $n$ at steps 5, 10, and 15\\[10pt]
\includegraphics[width=0.45\columnwidth]{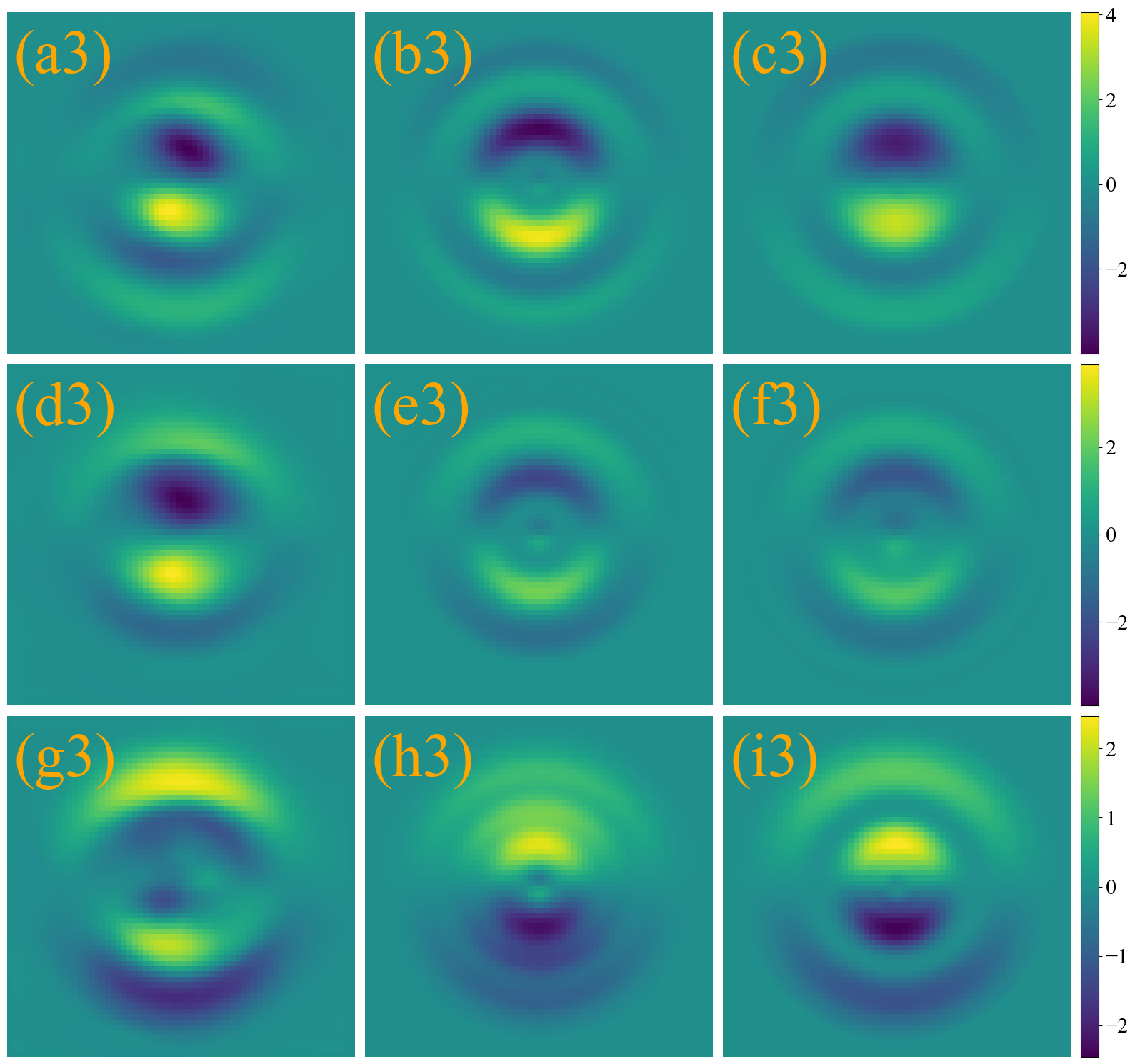} &
\includegraphics[width=0.45\columnwidth]{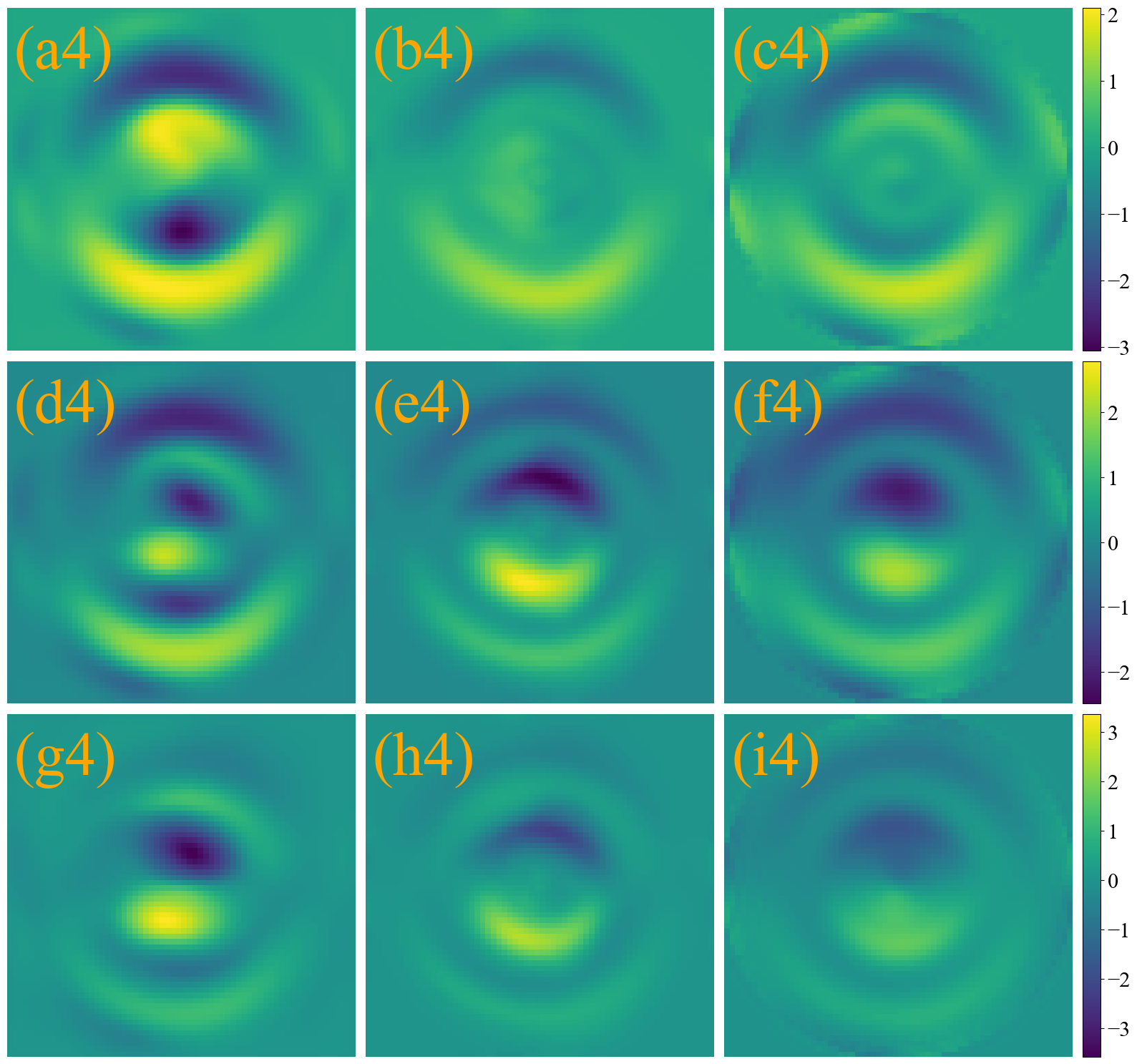} \\
(c) $E_z$ at steps 25, 30, and 35&
(d) $n$ at steps 25, 30, and 35
\end{tabular}
\caption{Comparison of predictions from the ConvLSTM and ExpNODE (as well as ExpNODE-p) models against the ground truth generated by MEGA code. The top row corresponds to the 20-step prediction horizon, while the bottom row corresponds to the 40-step prediction horizon. Within each subfigure, the left panel($a_j, d_j, g_j$) shows ConvLSTM predictions, the middle panel ($b_j, e_j, h_j$)shows the ExpNODE-p predictions, and the right panel ($c_j, f_j, i_j$) shows the ground truth.}
\label{fig:predictions}
\end{figure}

\subsection{Visual Comparison of Predictions}

To further illustrate the performance of the models, we provide visual comparisons of the predictions generated by the ConvLSTM and the ExpNODE-p models against the ground truth data. The predictions are from time steps 1,080 to 1,120, covering both the 20-step and 40-step prediction horizons.

From Figure~\ref{fig:predictions}, we observe that in short-term predictions, the ExpNODE-p model closely matches the ground truth, capturing detailed features and nonlinearities. Notably, the model accurately represents the asymmetry in the $\rho$ variable along the $z$-direction, which is a critical feature of the EGAM instability. In contrast, the ConvLSTM predictions lack fine details and degrade over time, showing increasing discrepancies with the ground truth at later time steps.

For long-range predictions (40-step horizon), the ExpNODE-p model performs excellently, and the quality of its predictions remains consistent throughout the prediction horizon. On the other hand, the ConvLSTM model provides limited information and fails to capture the essential dynamics, demonstrating its limitations in modeling complex, long-term spatiotemporal dependencies without incorporating physical constraints.

\section{Discussion}\label{sec4}

\subsection{Generalization and Applicability}

Our results imply that embedding physical information not only improves model performance on the immediate future beyond the data set it was trained on but also enables the model to generalize to new initial conditions outside the training range. To assess this, we evaluated the model's performance on data outside the training steps, specifically on sequences both before and after the training range.
For the 20-step prediction model, we assessed its performance on predicting time steps 820 to 840 (pre-training) and 1,085 to 1,105 (post-training). For the 40-step prediction model, evaluation focused on its ability to predict time steps 815 to 855 (pre-training) and 1,085 to 1,125 (post-training). The results are summarized in Table~\ref{tab:model_comparison_gen}.

The results show that the ExpNODE-p model performs comparably, and in some cases even better, on sequences outside the training set. This demonstrates the model's ability to generalize and accurately predict dynamics in regions where it was not trained. The slightly higher test losses in some cases are expected due to the differences in initial conditions, but the overall performance remains robust.

This ability to generalize is particularly important in the context of modeling plasma instabilities, where initial conditions can vary significantly. By learning the underlying physical constraints of the data rather than merely fitting to the training data patterns, the ExpNODE-p model exhibits enhanced predictive capabilities.
\begin{table}
    \centering
            \caption{Generalization performance of the ExpNODE-p model for 20-step and 40-step predictions on sequences outside the training set.}
    \label{tab:model_comparison_gen}
    \begin{tabular}{lcccc}
        \toprule
        & \multicolumn{2}{c}{\textbf{20-step Prediction}} & \multicolumn{2}{c}{\textbf{40-step Prediction}} \\
        \cmidrule(lr){2-3} \cmidrule(lr){4-5}
        \textbf{Test Range} & \textbf{Test Loss} & \textbf{Time Steps} & \textbf{Test Loss} & \textbf{Time Steps} \\
        \midrule
        Pre-training & 0.01970 & 820 to 840 & 0.1377 & 815 to 855 \\
        Test set & 0.0219 &  1,080 to 1,100 & 0.08772 &  1,080 to 1,120  \\ 
        Post-training & 0.05304 & 1,085 to 1,105 & 0.07798 & 1,085 to 1,125 \\
        \bottomrule
    \end{tabular}
\end{table}
\subsection{Implications and Applications}

The simplification methods and modeling approach presented in this work, while applied specifically to the EGAM instability, hold broad applicability across a wide range of plasma physics problems. The underlying principles of simplifying the MHD equations by leveraging physical insights, symmetry considerations, and perturbative analysis are fundamental techniques that can be extended to other plasma phenomena.

Moreover, the computational efficiency achieved by our model, even when trained on a relatively modest GPU (NVIDIA RTX 2060 Max-Q), suggests that with more powerful hardware, such as an RTX 4090 with significantly greater computing resources, the training time could be reduced substantially. This reduction in training time can enhance the efficiency of model testing and hyperparameter tuning, which are iterative and time-consuming processes. Consequently, it facilitates quicker deployment in practical applications, such as real-time prediction and control of plasma instabilities in fusion devices.
\subsection{Applicability to Other Plasma Phenomena and Extension to Machine Learning Models}

The core simplifications made in our method involve assuming dominance of certain terms in the governing equations based on physics.
These simplifications obviously improved the performance of the neutral network.
Such approximations are not unique to EGAM studies but are common in the modeling of various plasma waves and instabilities. For example, in the analyzing of Alfv\'en waves, similar assumptions are made to reduce the complexity of the equations and focus on the dominant physics.
Furthermore, the method of exploiting symmetries in the system to simplify the equations is widely applicable. In our case, we utilized the symmetry properties of the electric field and the plasma equilibrium to reduce the two-dimensional problem into a one-dimensional profile through the definition of a modified poloidal electric field \( E_{\text{pol, mod}}^{1D}(\rho) \). This approach can be extended to any plasma system exhibiting similar symmetries, such as quasi-symmetric stellarators, where certain spatial or temporal symmetries can be leveraged to simplify the analysis.

Integrating these simplifications with machine learning models enhances their ability to learn meaningful representations of the data, leading to improved performance. This methodology can be applied to other areas of plasma physics where machine learning models are employed, such as turbulence modeling, magnetic reconnection studies, or transport phenomena.

\subsection{Future Work and Improvements}

While our model demonstrates significant improvements over traditional methods, there are areas for further enhancement. As indicated in the training section, the output of the model requires Gaussian filtering as a post-processing step to smooth the rough edges resulting from numerical decoding. This is mainly due to the shape of the latent space $L(\rho,t)$. When transforming the latent space back into the original representation, numerical edges may occur, indicating that Fourier functions may not be the optimal choice for the kernel. Gaussian mixtures or other basis functions may be more promising candidates. Incorporating such functions could improve the smoothness of the output and reduce the need for post-processing.

Additionally, integrating the Gaussian filtering directly into the model training process, rather than applying it post hoc, could further reduce errors and enhance performance.
With access to more powerful computing resources, extensive hyperparameter tuning and model architecture exploration could lead to further improvements.
Moreover, expanding the model to handle more complex scenarios, such as three-dimensional data or additional physical processes, would increase its applicability. Collaborations with experimentalists and theorists could facilitate the incorporation of more detailed physical knowledge into the model, enhancing its predictive capabilities.

\section{Conclusion}\label{sec5}

We developed the ExpNODE-p framework, a modified neural ODE that integrates simplified physical laws into neural networks to model EGAMs in magnetically confined plasmas. By utilizing the dominant physical equations and symmetry considerations, essential EGAM dynamics were retained while reducing computational complexity.
The ExpNODE-p effectively captures the characteristics of EGAM profiles even in the nonlinear saturated stage and directly embeds physical constraints, enhancing prediction accuracy, efficiency, and generalizability. Evaluations using data from MEGA simulations demonstrate that ExpNODE-p outperforms models such as ConvLSTM and ExpNODE, achieving higher accuracy, faster convergence, and strong generalization beyond the training data range.

Our contributions are:

\textbf{Modification of neural ODE framework}: Designed the ExpNODE-p architecture that integrates physical laws into neural networks for improved modeling of complex plasma phenomena. The conventional neural ODE network is well improved by considering the the EGAM physics and symmetry.

\textbf{Demonstration of Improved Performance}: Showed that embedding physical information enhances predictive capabilities and generalization in machine learning models for plasma physics. 

This approach offers a promising pathway for tackling complex problems in plasma physics and related fields. Future work includes extending the ExpNODE-p framework to incorporate more detailed physical effects, applying it to other MHD instabilities and plasma phenomena, and integrating real experimental data to enhance its practical applicability in fusion devices.

In summary, the ExpNODE-p framework advances plasma dynamics modeling by combining physics-based modeling with machine learning, contributing to more accurate and efficient tools in plasma physics research and supporting the goal of achieving sustainable nuclear fusion energy.

~

\textbf{Acknowledgments}
The training data were generated on the ``Plasma Simulator'' (NEC SX-Aurora TSUBASA) of National Institute for Fusion Science (NIFS) with the support and under the auspices of the NIFS Collaboration Research program (NIFS22KIST025). This work was partially supported by the NINS program of Promoting Research by Networking among Institutions (Grant Number 01422301). The authors thank Prof. Y. Todo of NIFS for his sincere help and fruitful discussion on the training data generation and EGAM physics understanding.

\nocite{*}
\bibliographystyle{unsrt}
\bibliography{wh_ver2_pop}

\end{document}